\documentstyle[prl,aps,axodraw,floats]{revtex}

\def\bi{\bibitem}

\def\beb{\begin{thebibliography}{999}}
\def\eeb{\end{thebibliography}}
\def\bei{\begin{itemize}}
\def\eei{\end{itemize}}
\def\bef{\begin{figure}}
\def\eef{\end{figure}}
\def\ben{\begin{enumerate}}
\def\een{\end{enumerate}}
\def\beq{\begin{equation}}
\def\eeq{\end{equation}}
\def\ber{\begin{eqnarray}}
\def\eer{\end{eqnarray}}

\begin{document}
\draft
\twocolumn[
\hsize\textwidth\columnwidth\hsize\csname@twocolumnfalse\endcsname
\title{Neutrino Interactions in a Magnetized Medium}
\author{Kaushik Bhattacharya$^1$, Avijit K. Ganguly$^1$, Sushan Konar$^2$}
\address{ $^1$ Saha Institute of Nuclear Physics, 1/AF, Bidhan-Nagar, Calcutta 700064, India \\
$^2$IUCAA, Post Bag 4, Ganeshkhind, Pune 411007, India\\
e-mail : kaushikb@theory.saha.ernet.in, avijit@theory.saha.ernet.in,sushan@iucaa.ernet.in} 
\maketitle
\begin{abstract}
Neutrino-photon processes, forbidden in vacuum, can take place in presence of a thermal medium or an external 
electro-magnetic field, mediated by the corresponding charged leptons (real or virtual). The effect of a medium 
or an electromagnetic field is two-fold - to induce an effective $\nu-\gamma$ vertex and to modify the dispersion 
relations of all the particles involved to render the processes kinematically viable. It has already been noted 
that in presence of a thermal medium such an electromagnetic interaction translates into the neutrino acquiring 
a small effective charge. In this work, we extend this concept to the case of a thermal medium in presence of an 
external magnetic field and calculate the effective charge of a neutrino in the limit of a weak magnetic field. 
We find that the effective charge is direction dependent which is a direct effect of magnetic field breaking the 
isotropy of the space.                                                              
\end{abstract}
\narrowtext
\bigskip
]
\section{Introduction}
\label{intro}
 
Processes involving neutrinos and photons are of great importance in astrophysics and cosmology~\cite{raff}.
In particular, reactions which are forbidden (or are highly suppressed) in vacuum, notably plasmon decay
($\gamma \rightarrow \nu \nu$) or the Cherenkov process ($\nu \rightarrow \nu \gamma$) and the cross-processes
(e.g, $\gamma \gamma \to \nu \bar{\nu}$, $\nu \bar{\nu} \to e^+ e^-$ etc.) play a significant role in regions
pervaded by dense plasma and/or large-scale external magnetic fields. White Dwarfs, Neutron Stars or the final
phases of stellar evolution (Supernovae) are particular examples where such processes become important by virtue
of the large material density and the presence of strong magnetic fields. Prompted by these objectives we calculate 
the effective neutrino charge in an external magnetic field in presence of a thermal medium. 

It has already been shown that the $\nu-\gamma$ interaction in presence of a thermal medium induces a small 
effective charge to the neutrino and that the neutrino electromagnetic vertex is related to the photon self-energy 
in the medium~\cite{pal2,alth}. We re-investigate this problem considering not only a thermal medium but also an 
external magnetic field for a neutrino coupled to a dynamical photon having $q_0=0$ and $|\vec{q}| \to 0 $. This 
calculation is pertinent, for example, in the case of a type-II supernovae collapse~\cite{colp}. It is conjectured 
that the neutrinos, produced deep inside the stellar core, deposit some fraction of their energy to the surrounding 
medium through different kinds of electromagnetic interactions, e.g, $\nu \to \nu \gamma$, 
$\nu \bar{\nu} \to e^{+} e^{-}$~\cite{expl}. But it is important to note that the amplitudes of such processes are 
proportional to $G^2_F$ and the amount of energy transferred to the mantle of the proto-neutron star is barely sufficient 
to blow the stellar envelope out. Recently in a series of papers, it has been argued that the freely streaming neutrinos 
from the supernova core interact with the non-relativistic electrons present in the outer part of the core through collective 
interactions (known as the `two-stream instability' in the context of plasma physics) and is responsible for blowing up the 
mantle of the supernova progenitor~\cite{shkl}. 

In this work, we show that the effective charge acquired by a neutrino in a magnetised medium is, in fact, {\it direction 
dependent} which should affect these processes significantly. Though it should be noted here that except for in the very 
early universe or in the recently discovered `magnetars' (newly born neutron stars with magnetic fields in excess of 
$10^{15}$~Gauss)~\cite{kouv} or perhaps in the central region of core-collapse supernovae, the magnetic fields are smaller 
than the QED limit ($eB < m_e^2$ i.e, $B \le 10^{13}$~Gauss). This allows for a weak-field treatment of the plasma processes 
relevant to almost all physical situations. Moreover, this treatment is also valid for compact astrophysical objects (viz.
white dwarfs or neutron stars) for which the Landau level spacings are quite small compared to the electron Fermi
energy~\cite{chan}. This ensures that the magnetic field does not introduce any spatial anisotropy in the collective
plasma behaviour. 
 
In the standard model, the above mentioned $\nu-\gamma$ processes appear at the one-loop level. They do not 
occur in vacuum because they are kinematically forbidden and also because the neutrinos do not couple to the 
photons at the tree-level. In the presence of a medium or a magnetic field, it is the charged particle running 
in the loop which, when integrated out, confers its electromagnetic properties to the neutrino. Therefore, 
processes involving two neutrinos and one photon which are forbidden in vacuum can become important in the 
presence of a medium and/or external fields~\cite{hari,pal1,iorf}. These charged particles could be the electrons 
of the background thermal medium or the virtual electrons and positrons in presence of an external magnetic field 
or both. The processes also become kinematically allowed since the photons acquire an effective mass in a thermal 
medium. This, for example, opens up the phase space for the Cherenkov process $\nu \rightarrow \nu \gamma$ 
~\cite{pal1,hame,orae}. The presence of a magnetic field would also modify the photon dispersion relation and then 
the Cherenkov condition would be satisfied for significant ranges of photon frequencies~\cite{shai,gsh1,gsh2}. A 
thermal medium and an external magnetic field, thus, fulfill the dual purpose of inducing an effective neutrino-photon 
vertex and of modifying the photon dispersion relation such that the phase-space for various neutrino-photon processes 
is opened up(see~\cite{iorf} and references therein for a detailed review). 
                                                                                                            
The organization of the paper is as follows. In section-\ref{form} we discuss the basic formalism for calculating the
effective charge of a neutrino. Section-\ref{calc} contains the details of the calculation of the 1-loop diagram in 
presence of a magnetised medium. In section-\ref{tensor} we discuss the generic form of the neutrino effective charge
in different background conditions. Finally, in section-\ref{nech} we present the expression for the effective charge 
of a neutrino. And we conclude with a discussion on the possible implications of our result in section-\ref{concl}. 
 
\section{Formalism}
\label{form}
 
The off-shell electromagnetic vertex function $\Gamma_{\lambda}$ is defined in such a way that, for on-shell
neutrinos, the $\nu \nu \gamma$ amplitude is given by:
\beq
{\cal M} = - i \bar{u}(k') \Gamma_{\lambda} u(k) A^{\lambda}(q),
\eeq
where, $q,k,k'$ are the momentum carried by the photon and the neutrinos respectively and $q=k-k'$. Here, $u(k)$ is
the neutrino wave-function and $A^{\mu}$ stands for the electromagnetic vector potential. In general, $\Gamma_{\lambda}$ 
would depend on $k$, $q$, the characteristic of the medium and the external electromagnetic field.  We shall, in this work, 
consider neutrino momenta that are small compared to the masses of the W and Z bosons. We can, therefore, neglect the 
momentum dependence in the W and Z propagators, which is justified if we are performing a calculation to the leading order 
in the Fermi constant, $G_F$. In this limit four-fermion interaction is given by the following effective Lagrangian:
\beq
{\cal L}_{\rm eff} = \frac{1}{\sqrt{2}} G_F {\overline \nu} \gamma^{\mu} (1 + \gamma_5) \nu {\overline l_\nu}
                     \gamma_{\mu} (g_{\rm V} + g_{\rm A} \gamma_5) l_\nu \,,
\eeq
where, $\nu$ and $l_\nu$ are the neutrino and the corresponding lepton field respectively. For electron neutrinos,
%
\ber
g_{\rm V} &=& 1 - (1 - 4 \sin^2 \theta_{\rm W})/2, \\
g_{\rm A} &=& 1 - 1/2;
\eer
where the first terms in $g_{\rm V}$ and $g_{\rm A}$ are the contributions from the W exchange diagram and the second
one from the Z exchange diagram. Then the amplitude effectively reduces to that of a purely photonic case with one of the
photons replaced by the neutrino current, as seen in the diagram in fig.~\ref{f:cher}. Therefore, $\Gamma_{\lambda}$ is given
by:
\beq
\Gamma_{\mu} = - \frac{1}{\sqrt{2}e} G_F \gamma^{\nu} (1 + \gamma_5) \,(g_{\rm V} \Pi_{\mu \nu} + g_{\rm A} \Pi_{\mu \nu}^5) \,,
\eeq
where, $\Pi_{\mu \nu}^5$ represents the vector-axial vector coupling and $\Pi_{\mu \nu}$ is the polarisation tensor arising
from the diagram in fig.~\ref{f:polr}. In an earlier paper~\cite{frd1} (paper-I henceforth) we have analysed the structure 
of $\Pi^{\mu \nu}$ and calculated the photon dispersion relation, in a background medium in presence of a uniform external 
magnetic field, in the weak-field limit by retaining terms up-to ${\cal O(B)}$, calculated at the 1-loop level. We shall use 
the results of paper-I here to obtain the total effective charge of the neutrinos under equivalent conditions. Because of 
the electromagnetic current conservation, for the polarisation tensor, we have the following gauge invariance condition:
\beq
q^{\mu} \Pi_{\mu \nu} = 0 = \Pi_{\mu \nu} q^{\nu}.
\eeq
Same is true for the photon vertex of fig.~\ref{f:cher} and we have
\beq
\Pi_{\mu \nu}^5 q^{\nu} = 0\,.
\label{gi_pi5}
\eeq

Therefore, the effective charge of the neutrinos is defined in terms of the vertex function by the following
relation~\cite{pal2}:
\beq
e_{\rm eff} = {1\over{2 k_0}} \, \bar{u}(k) \, \Gamma_0(q_0=0, {\bf q} \rightarrow 0) \, u(k) \,.
\eeq
For massless Weyl spinors this definition can be rendered into the form:
\beq
e_{\rm eff} = {1\over{2 k_0}} \, \mbox{tr} \left[\Gamma_0(q_0=0, {\bf q}\rightarrow 0) \, (1+\lambda \gamma^5) \, \rlap/k \right]
\label{nec1}
\eeq
where $\lambda = \pm 1$ is the helicity of the spinors.
 
It can be seen from Eq.(\ref{nec1}) that, in general, the effective neutrino charge depends on $\Pi_{\mu\nu}(q)$ as
well as on $\Pi^{5}_{\mu\nu}(q)$. Now, in a magnetised medium, the dispersive part of $\Pi_{\mu \nu}$ to liner order 
in ${\cal B}$ has the following form [paper-I]:
\beq
\Pi_{\mu \nu} \propto \varepsilon_{\mu \nu \alpha_{\parallel} \beta} q^{\beta} \,,
\eeq
where $\alpha_\parallel$ stands for either 0 or 3 (to be explained in detail in the next section). This evidently vanishes
in the limit $q_0 =0, {\bf q} \rightarrow 0$. Therefore, in a magnetised medium, the non-zero contribution to the effective 
charge of the neutrinos come solely from $\Pi^{5}_{\mu\nu}(q)$. In section-\ref{tensor} we shall discuss, from a more general
point of view, why the effective charge of a neutrino, in a magnetised medium, comes only from $\Pi^{5}_{\mu\nu}(q)$ to linear 
order in ${\cal B}$. Therefore, the effective charge of the neutrinos is given by:
\ber
e_{\rm eff} &=& - {1\over{2 k_0}} \, \frac{G_F}{\sqrt{2}e} \, g_{\rm A} \Pi_{\mu 0}^5(q_0=0, {\bf q}\rightarrow 0) \nonumber \\
&& \times \, \mbox{tr} \left\{\gamma^{\mu} (1 + \gamma_5) \, (1+\lambda \gamma^5) \, \rlap/k \right\} \,.
\label{nec2}
\eer
%

\section{calculation of the 1-loop diagram}
\label{calc}
 
\subsection{The Propagator}
 
\bef
\begin{center}
\begin{picture}(150,40)(0,-35)
\Photon(0,0)(40,0){2}{4}
\Text(20,5)[b]{$q\rightarrow$}
\Text(75,30)[b]{$p+q\equiv p'$}
\Text(125,-17)[b]{$k$}
\Text(125,15)[b]{$k'$}
\Text(47,0)[]{$\nu$}
\Text(102,0)[]{$\mu$}
\Text(75,-30)[t]{$p$}
\ArrowLine(110,0)(160,30)
\ArrowLine(160,-30)(110,0)
\SetWidth{1.2}
\Oval(75,0)(25,35)(0)
\ArrowLine(74,25)(76,25)
\ArrowLine(76,-25)(74,-25)
\end{picture}
\end{center}
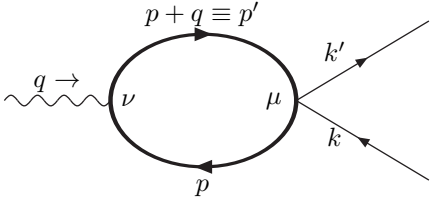
\caption[]{One-loop diagram for the effective electromagnetic vertex of the neutrino in the limit of infinitely
heavy W and Z masses.}\label{f:cher}
\eef
\bef
\begin{center}
\begin{picture}(150,50)(0,-25)
\Photon(0,0)(40,0){2}{4}
\Text(20,5)[b]{$q\rightarrow$}
\Photon(110,0)(150,0){2}{4}
\Text(130,5)[b]{$q\rightarrow$}                                                           
\Text(75,30)[b]{$p+q\equiv p'$}
\Text(75,-30)[t]{$p$}
\SetWidth{1.2}
\Oval(75,0)(25,35)(0)
\ArrowLine(74,25)(76,25)
\ArrowLine(76,-25)(74,-25)
\end{picture}
\end{center}
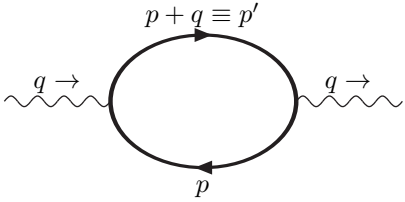
\caption[]{One-loop diagram for the polarisation tensor.}\label{f:polr}
\eef
 
Since we investigate the case of a purely magnetic field, it can be taken in the $z$-direction without any further loss 
of generality. We denote the magnitude of this field by $\cal B$. Ignoring at first the presence of the medium, the electron 
propagator in such a field can be written down following Schwinger's approach~\cite{schw,tsai,ditt}:
\beq
i S_B^V(p) = \int_0^\infty ds \, e^{\Phi(p,s)} \, C(p,s) \,,
\label{SV}
\eeq
where $\Phi$ and $C$ are defined below. To write these in a compact notation, we decompose the metric tensor into two parts:
\beq
g_{\mu \nu} = g^{\parallel}_{\mu \nu} + g^{\perp}_{\mu \nu} \,,
\eeq
where
\ber
g^{\parallel}_{\mu \nu} &=& diag(1,0,0-1) \, \nonumber \\
g^{\perp}_{\mu \nu} &=& diag(0,-1,-1,0)\,.
\eer
This allows us to use the following definitions,
\ber
p_\parallel^2 &=& p_0^2 - p_3^2 \, \\
p_\perp^2 &=& p_1^2 + p_2^2 \,.
\eer
Using these notations we can write:
\ber
\Phi(p,s) &\equiv& 
          is \left( p_\parallel^2 - {\tan (e{\cal B}s) \over e{\cal B}s} \, p_\perp^2 - m^2 \right) - \epsilon |s| \,,
\label{Phi} \\
C(p,s) &\equiv& {e^{ie{\cal B}s\sigma\!_z} \over \cos(e{\cal B}s)} 
       \, \left( \rlap/p_\parallel + \frac{e^{-ie{\cal B}s\sigma_z}} {\cos(e{\cal B}s)}\rlap/ p_\perp + m \right) \nonumber \\
       &=& \Big[ ( 1 + i\sigma_z \tan  e{\cal B}s ) (\rlap/p_\parallel + m ) + (\sec^2 e{\cal B}s) \rlap/ p_\perp \Big] \,,
\label{C}
\eer
where
\beq
\sigma_z = i\gamma_1 \gamma_2 = - \gamma_0 \gamma_3 \gamma_5 \,,
\label{sigz}
\eeq
and we have used,
\beq
e^{ie{\cal B}s\sigma_z} = \cos \; e{\cal B}s + i\sigma_z \sin \; e{\cal B}s \,.
\eeq
Of course in the range of integration indicated in Eq.~(\ref{SV}) $s$ is never negative and hence $|s|$ equals $s$.
It should be mentioned here that we follow the notation adopted in paper-I to ensure continuity. In the presence of
a background medium, the above propagator is modified to~\cite{elmf}:        
\beq
iS(p) = iS_B^V(p) + S_B^\eta(p) \,,
\label{fullprop}
\eeq
where
\beq
S_B^\eta(p) \equiv - \eta_F(p) \left[ iS_B^V(p) - i\overline S_B^V(p) \right] \,,
\eeq
and 
\beq
\overline S_B^V(p) \equiv \gamma_0 S^{V \dagger}_B(p) \gamma_0 \,,
\label{Sbar}
\eeq
for a fermion propagator, such that
\beq
S_B^\eta(p) = - \eta_F(p) \int_{-\infty}^\infty ds\; e^{\Phi(p,s)} C(p,s) \,.
\label{Seta}
\eeq
And $\eta_F(p)$ contains the distribution function for the fermions and the anti-fermions:
\ber
\eta_F(p) &=& \Theta(p\cdot u) f_F(p,\mu,\beta) \nonumber \\
&+& \Theta(-p\cdot u) f_F(-p,-\mu,\beta) \,.
\label{eta}
\eer
Here, $f_F$ denotes the Fermi-Dirac distribution function:
\beq
f_F(p,\mu,\beta) = {1\over e^{\beta(p\cdot u - \mu)} + 1} \,,
\eeq
and $\Theta$ is the step function given by:
\ber
\Theta(x) &=& 1, \; \mbox{for $x > 0$} \,, \nonumber \\
&=& 0, \; \mbox{for $x < 0$} \,. \nonumber                
\eer
%
\subsection{Identifying the Relevant Terms}
 
The amplitude of the 1-loop diagram of fig.~\ref{f:cher} can be written as:
\beq
i \Pi^5_{\mu\nu}(q, \beta) = - \int \frac{d^4p}{(2\pi)^4} (ie)^2 \; \mbox{tr}\,
\left[\gamma_\mu \, \gamma_5 \, iS(p) \gamma_\nu \, iS(p')\right] \,,
\eeq
where, for the sake of notational simplicity, we have used
\beq
p' = p+q \,.
\label{p'}
\eeq
The minus sign on the right side is for a closed fermion loop and $S(p)$ is the
propagator given by Eq.~(\ref{fullprop}). This implies:
\beq
\Pi^5_{\mu\nu}(q, \beta) = -ie^2 \int \frac{d^4p}{(2\pi)^4} \;
\mbox{tr}\, \left[\gamma_\mu \, \gamma_5 \, iS(p) \gamma_\nu \, iS(p')\right] \,.
\label{1loopampl}
\eeq
Now using Eq.(\ref{Seta}) we have the terms containing the effects of medium and the external field (non-absorptive)
\ber
\Pi^5_{\mu\nu}(q, \beta) &=& -ie^2 \int \frac{d^4p}{(2\pi)^4} \;
\mbox{tr}\, \Big[\gamma_\mu \, \gamma_5 \, S_B^\eta(p) \, \gamma_\nu \, iS_B^V(p') \nonumber \\
&& + \gamma_\mu \, \, \gamma_5 \, iS_B^V(p) \, \gamma_\nu \, S_B^\eta (p') \Big] \,.
\label{SS'terms}
\eer
Substituting $p$ by $-p'$ in the second term and using the cyclic property of traces, we can write Eq.~(\ref{SS'terms}) as
\ber
\Pi^5_{\mu\nu}(q, \beta) &=& -ie^2 \int \frac{d^4p}{(2\pi)^4} \;
\mbox{tr}\, \Big[\gamma_\mu \, \gamma_5 \, S_B^\eta(p) \, \gamma_\nu \, iS_B^V(p') \nonumber \\     
&& + \gamma_\nu \, S_B^\eta(-p) \, \gamma_\mu \, \gamma_5 \, iS_B^V (-p') \Big] \,.
\label{SS'terms2}
\eer
Using now the form of the propagators from Eqs.(\ref{SV}) and (\ref{Seta}), we obtain
\ber
\Pi^5_{\mu\nu}(q, \beta)
&=& ie^2 \int \frac{d^4p}{(2\pi)^4} \int_{-\infty}^\infty ds \; e^{\Phi(p,s)} \int_0^\infty ds' \; e^{\Phi(p',s')} \nonumber\\*
&& \times \, {\cal G}_{\mu \nu}(p,p',s,s',B) \,,
\label{Pi5}
\eer
with,
\ber
{\cal G}_{\mu \nu}
&=& \eta_F(-p) \, \mbox{tr} \, \Big[\gamma_\nu \, C(-p,s) \, \gamma_\mu \, \gamma_5 \, C(-p',s')\Big] \nonumber \\
&+& \eta_F(p) \, \mbox{tr} \, \Big[\gamma_\mu \, \gamma_5 \, C(p,s) \, \gamma_\nu \, C(p',s')\Big].
\eer
It should be mentioned here that the effective charge of the neutrinos come from the dispersive part of the axial polarisation
tensor. Therefore, we work with the real part of the 11-component of the axial polarisation tensor throughout and for notational
simplicity suppress the 11-index everywhere.
%
\subsection{Extracting the Gauge Invariant Piece}
 
{\bf $\Pi^5_{\mu\nu}(k, \beta)$ in odd powers of ${\cal B}$} - Notice that the phase factors appearing in Eq.~(\ref{Pi5})
are even in $\cal B$. Thus, we need consider only the odd terms from the traces. Performing the traces the gauge invariant
expression, odd in powers of ${\cal B}$, comes out to be:
\ber
\Pi^5_{\mu\nu}(q, \beta) &=& - 4e^2 \, \int \frac{d^4p}{(2\pi)^4} \eta_+(p) \int_{-\infty}^\infty ds \, e^{\Phi(p,s)} \nonumber \\
&\times& \int_{-\infty}^\infty ds' \, e^{\Phi(p',s')} \; R_{\mu\nu} \,;
\label{pi5_odd}
\eer
where
\beq
\eta_+(p) = \eta_F(p) + \eta_F(-p)                   
\eeq
and
\ber
{\cal R}_{\mu\nu} &=& - \varepsilon_{\mu \nu 1 2} \, \frac{\sec^2 e{\cal B}s \tan^2 e{\cal B}s'}{\tan e{\cal B}s + \tan e{\cal B}s'}
                      q_{\perp}^2 \nonumber\\
                  &-& \varepsilon_{\mu \nu 1 2} \, (q\cdot p) \, (\tan e{\cal B}s + \tan e{\cal B}s') \nonumber\\
                  &+& 2\varepsilon_{\mu 1 2 \alpha_\parallel} \, (p'_{\nu_\parallel} p^{\alpha_\parallel} \tan e{\cal B}s 
                       + p_{\nu_\parallel} p'^{\alpha_\parallel} \tan e{\cal B}s') \nonumber\\
                  &+& g_{\mu\alpha_\parallel} q_{\nu_\perp} 
                      p^{\widetilde \alpha_\parallel} \, (\tan e{\cal B}s - \tan e{\cal B}s') \nonumber\\
                  &-& g_{\mu\alpha_\parallel} q_{\nu_\perp} q^{\widetilde \alpha_\parallel}
                      \, \frac{\sec^2 e{\cal B}s \, \tan^2 e{\cal B}s'}{\tan e{\cal B}s + \tan e{\cal B}s'} \nonumber\\
                  &+& \left\{g_{\mu\nu} (p\cdot \widetilde q)_\parallel 
                            + g_{\nu \alpha_\parallel} p^{\widetilde \alpha_\parallel} q_{\mu_\perp} \right\} 
                      \, (\tan e{\cal B}s - \tan e{\cal B}s') \nonumber\\
                  &+& g_{\nu \alpha_\parallel} q^{\widetilde\alpha_\parallel} p_{\mu_\perp} \sec^2 e{\cal B}s \, \tan e{\cal B}s' \,.
\label{rmunu}
\eer
In writing this expression, we have used the notation $p^{\widetilde\alpha_\parallel}$, for example. This signifies 
that $\widetilde\alpha_{\parallel}$ can take only the `parallel' indices, i.e., 0 and 3, and is moreover different from 
the index $\alpha$ appearing elsewhere in the expression. We perform the calculations in the rest frame of the medium where
$p\cdot u=p_0$. Thus the distribution function does not depend on the spatial components of $p$ and is given simply by
$\eta_+(p_0)$. 
%
\section{Tensorial Structures of $\Pi_{\mu \nu}$ and $\Pi^5_{\mu \nu}$}
\label{tensor}

It can be seen from Eq.(\ref{nec1}) that, in general, the effective neutrino charge depends on $\Pi_{\mu\nu}(q)$ as
well as on $\Pi^{5}_{\mu\nu}(q)$. Now, in vacuum we have,
\beq
\Pi_{\mu\nu}(q) = (g_{\mu\nu} q^2 - q_{\mu}q_{\nu}) \,\Pi(q^2) \,,
\eeq
where $\Pi(k^2)$ vanishes for $q_0=0, \bar{q}\rightarrow 0$. The other contribution to the effective charge, coming from 
$\Pi^{5}_{\mu\nu}(k)$, also vanishes in vacuum. This can be understood from the general form factor analysis. We should be
able to express $\Pi^{5}_{\mu\nu}(k)$, in vacuum, in terms of $g_{\mu\nu}$, $\epsilon_{\mu\nu\lambda\sigma}$ and $q_{\lambda}$. 
The parity structure of the theory forbids the appearance of $g_{\mu\nu}$. Therefore, the only possible combination, to obtain 
a second rank tensor is, $\epsilon_{\mu \nu \lambda \sigma} q_\lambda q_\sigma$. Since, this is identically zero, there can be 
no effective charge of a neutrino in vacuum. 

On the other hand, in the presence of a medium the polarisation tensor can be expanded in terms of the form factors as 
follows~\cite{pal1}: 
\ber
\Pi_{\mu\nu}(k) = \Pi_T\,T_{\mu\nu} + \Pi_L\,L_{\mu\nu} + \Pi_P\,P_{\mu\nu} \,,
\eer
where
\ber
T_{\mu\nu} &=& {\widetilde g}_{\mu\nu} - L_{\mu\nu} \, \\
L_{\mu\nu} &=&\frac{{\widetilde u}_{\mu}{\widetilde u}_{\nu}}{{\widetilde u}^2} \, \\
P_{\mu\nu} &=&\frac{i}{{\cal Q}} \varepsilon_{\mu \nu \alpha \beta}q^{\alpha} u^{\beta} \,
\eer
and,
\ber
{\widetilde g}_{\mu\nu} &=& g_{\mu\nu} - \frac{q_{\mu}q_{\nu}}{k^2} \, \\
{\widetilde u}_{\mu} &=& {\widetilde g}_{\mu\rho} u^{\rho} \, \\
{\cal Q} &=& \sqrt{(q\cdot u)^2 - q^2} \,
\eer
in the rest frame of the medium where $u^{\mu}=(1,0,0,0)$. It is easy to see that neither the longitudinal projection 
$L_{\mu \nu}$ or $\Pi_L$ is non-zero in the limit $q_0=0, \bar{q}\rightarrow 0$. This then provides a non-zero contribution 
to the effective charge of the neutrino. And in ~\cite{pal2} it has been shown that the non-zero contribution to the effective
charge, in presence of a thermal medium, comes only from the $\Pi_{\mu\nu}(q)$ part. The tensor structure of $\Pi^{5}_{\mu\nu}$ 
in a medium is of the form $\varepsilon_{\mu\nu\alpha\beta} \, q^{\alpha} u^{\beta}$ and does not contribute to the zeroth 
component of $\Gamma_{\nu}$. For a more physical understanding of the appearance of the effective charge of the neutrinos, 
in a medium, see~\cite{pal2}.

Now, at the 1-loop level $\Pi_{\mu \nu}$ is invariant under charge conjugation, i.e., if we calculate the vacuum polarisation
in a medium with a certain background field, it should be the same as that obtained in a charge-conjugated medium with an
opposite background field. This means that, in the polarization tensor, the terms are either even in the background field 
or even in $\mu$ or odd in both. Therefore, in absence of a medium (which can be thought of as containing $\mu^0$), the terms
containing odd powers of the background field should vanish (see paper-I for a discussion). Therefore, the contribution to the 
effective charge of a neutrino, in presence of a background magnetic field in vacuum comes only from $\Pi^{5}_{\mu\nu}(q)$, 
to linear order in ${\cal B}$. Now, $\Pi^5_{\mu\nu}(q)$ in a magnetised vacuum is given by \cite{hari}:
\ber
\Pi^5_{\mu\nu}(q) &=& \frac{i}{(4\pi)^2}\int^{\infty}_{0} ds \int^{1}_{-1}dv/2 e^{-is \chi} \nonumber \\
                  && \Big\{(1- v^2)q^{\mu}_{\parallel} e\tilde{Fq}_{\nu} \nonumber \\
                  && + R \left(-q^{2}_{\perp} e \tilde{F}^{\mu\nu} 
                     + q^{\mu}_{\perp} e\tilde{Fq}_{\nu}+q^{\nu}_{\perp} e\tilde{Fq}_{\mu} \right) \Big\} \,,
\label{hd}
\eer
with, 
\ber
\chi &=& m^2 + \frac{(1- v^2)}{4} q^2_{\parallel} + \frac{\cos vz- cos z}{2z \sin z} \\
R &=& \frac{1-v \sin vz \sin z - \cos vz \cos z}{{\sin}^2 z} \,,
\label{h}
\eer
where, $z=eBs$ and $\tilde{F}^{\mu\nu}=\frac{1}{2} \epsilon^{\mu\nu\lambda\sigma}F_{\lambda\sigma}$ (note that the metric used 
in Eq.~(\ref{hd}) is $g^{\mu\nu}= diag(-,+,+,+)$). It is evident that in the zero frequency and long wavelength limit
$\Pi^{5}_{\mu\nu}$ vanishes resulting in zero effective charge of a neutrino in magnetised vacuum.

The tensorial form of $\Pi_{\mu \nu}$ has been discussed in detail in paper-I. It can be seen from that discussion that
in a magnetised medium, the electromagnetic field always appears in the combination $u^{\mu} F_{\mu \nu}$ or $q^{\mu} F_{\mu \nu}$ 
in $\Pi_{\mu \nu}$, to linear order in the field strength. Since, we consider the case of a pure magnetic field
and a stationary medium the only terms that would survive would contain the combination $q^{\mu} F_{\mu \nu}$ (also borne out
by the explicit calculations of paper-I). Hence, in the zero frequency, long wavelength limit it vanishes leaving only 
$\Pi^5_{\mu \nu}$ to contribute to the effective charge of the neutrino. Now, in a magnetized medium $\Pi^5_{\mu \nu}$ can be 
written, in terms of general basis vectors available, as follows:
\ber
\Pi^{5}_{\mu\nu} &=& \epsilon_{\mu\nu 12} (q^2_{\perp} f_1 + q^2_{\parallel}f_2 + (q.u) f_3) \nonumber \\
                 &+& \epsilon_{\mu 12 \alpha_{\parallel}} (q_{\nu_{\parallel}} u^{\alpha_{\parallel}} f_3 
                     + u_{\nu_{\parallel}} u^{\alpha_{\parallel}} f_4 + q^{\alpha_{\parallel}} u_{\nu_{\parallel}} f_5  \nonumber \\
                 && + q_{\nu_{\parallel}}q^{\alpha_{\parallel}} f_6 + u_{\nu_{\parallel}} q^{\alpha_{\parallel}} f_7 
                    + q^{\alpha^{\parallel}} q_{\nu_{\parallel}} f_8) \nonumber \\
                &+& g_{\mu \alpha_{\parallel}} q_{\nu_{\perp}} (u^{\tilde{\alpha_{\parallel}}}f_9 
                    + q^{\tilde{\alpha_{\parallel}}} f_{10}) \nonumber \\
                &+& g_{\mu\nu} q_{\alpha_{\parallel}} q^{\tilde{\alpha}_{\parallel}} f_{11}
                    + g_{\mu\nu} u_{\alpha_{\parallel}} q^{\tilde{\alpha}_{\parallel}} f_{12}
                    + g_{\nu  \alpha_{\parallel}} u^{\tilde{\alpha_{\parallel}}} q_{\mu_{\perp}} f_{13} \nonumber \\
                && + \, g_{\nu  \alpha_{\parallel}} q^{\tilde{\alpha_{\parallel}}} q_{\mu_{\perp}} f_{14} 
                    + g_{\nu  \alpha_{\parallel}}  q^{\tilde{\alpha_{\parallel}}} q_{\mu_{\perp}} f_{15} \,,
\label{ff-api5}
\eer
where $f_i$s are the respective form factors. It can be easily seen that the terms proportional to the product of $u$'s 
are non-zero in the static long wavelength limit giving a finite contribution to the effective charge of a neutrino.

In this connection, it should be mentioned that the effective charge of the neutrinos bear a simple relation with the Debye 
screening length in the case of an unmagnetized plasma. As has been shown by \cite{pal2} the contribution to the effective 
charge comes only from $\Pi_L$ which corresponds to the Debye screening in the limit $q_0 = 0, {\bf q} \rightarrow 0$. In the 
case of a magnetised plasma $\Pi^5_{\mu \nu}$, in general, would have many more tensorial forms in it due to the presence of 
the electromagnetic tensor (paper-I0. Hence, there may not exist a simple correspondence between the Debye screening and the 
effective charge of a neutrino in a magnetised medium.
 
\section{Effective Charge of a Neutrino }
\label{nech}
 
It is evident from Eq.(\ref{nec2}) that to find the effective charge of the neutrino we need only to calculate
$\Pi^5_{\mu 0}$ in the limit $(q_0=0,{\bf q} \rightarrow 0)$. From Eq.(\ref{rmunu}) it can be seen that in this
limit the only surviving terms in $\Pi^5_{00}$ are the ones containing odd powers of $p_0$. Now, $\eta_+(p_0)$ is even
in $p_0$ and so is the exponent in the zero frequency limit. Hence, $p_0$ integration makes $\Pi^5_{00}$ vanish. We
also have,
\ber
R_{10} &=& p_3 q_1 (\tan e{\cal B}s - \tan e{\cal B}s') \nonumber \\
       &&  + q_3 p_1 \sec^2 e{\cal B}s \tan e{\cal B}s' \,,\\
R_{2 0} &=& p_3 q_2 (\tan e{\cal B}s - \tan e{\cal B}s') \nonumber\\
        && + q_3 p_2 \sec^2 e{\cal B}s \tan e{\cal B}s' \,.
\eer
It can be seen that after $p$ integration we shall have terms proportional to $q_3 q_1$ and $q_3 q_2$ in $\Pi^5_{1 0}$ and 
$\Pi^5_{2 0}$ respectively, as the integrals in $p_1$ and $p_2$ are Gaussian. In the limit ${\bf q}\rightarrow 0$ these terms
would vanish.  Therefore, in the relevant limit of vanishing external photon momenta only $\Pi^5_{3 0}$, given by,
\ber
&& \Pi^5_{3 0}(q_0 = 0, \vec q \rightarrow 0)  \nonumber \\
&=& \lim_{q_0=0,\vec{q}\rightarrow 0} 4 e^2 \int{d^4p\over{(2\pi)^4}}  
    \int^{\infty}_{-\infty} ds\, e^{\Phi(p,s)} \int^{\infty}_0 ds' e^{\Phi(p',s')}\nonumber\\
&\times& \eta_+(p_0) \left((q\cdot p)_\parallel - 2 p^2_0 \right) \, (\tan e{\cal B}s + \tan e{\cal B}s') 
\label{pi5k0}
\eer
has a non-zero contribution to the effective charge. It can be seen that in the above expression, except for the
exponents, the integrand if free of the perpendicular components of the momentum. Therefore, the perpendicular components of 
the loop momentum can be integrated out. Now, the exponential factors can be written as,
\ber
\Phi(p,s) + \Phi(p',s') = \Phi_\parallel + \Phi_\perp \,,
\eer
where,
\ber
\Phi_\parallel &=& is \, (p^2_\parallel - m^2)  + is' \, (p'^2_\parallel - m^2)  - \varepsilon \, |s| - \varepsilon \, |s'| \,, \\
\Phi_\perp &=& - \frac{i\tan e{\cal B}s}{e{\cal B}} \, p^2_\perp - \frac{i\tan e{\cal B}s'}{e{\cal B}} \, p'^2_\perp \,.
\label{phex}
\eer
Therefore, integration of the perpendicular part of the momentum gives us,
\ber
&& \int {d^2 p_\perp \over{(2 \pi})^2}\, e^{\Phi_\perp} \nonumber \\
&& = \exp\left( - \frac{i}{e{\cal B}} \frac{\tan e{\cal B}s \tan e{\cal B} s'}{\tan e{\cal B}s + \tan e{\cal B} s'} k_{\perp}^2\right)
   \, \int {d^2 p_\perp \over{(2 \pi})^2} \nonumber \\
&& \times \, \exp\left(-i \frac{\tan e{\cal B}s \tan e{\cal B} s'}{e{\cal B}}
   (p_\perp + \frac{\tan e{\cal B} s'}{\tan e{\cal B}s + \tan e{\cal B} s'} q_\perp)^2\right) \nonumber \\
&& =  - \frac{1}{4\pi}\frac{ie{\cal B}}{\tan e{\cal B}s + \tan e{\cal B}s'} \,;
\label{pri}
\eer
where we have neglected terms up-to ${\cal O}(q_\perp^2)$ since, to calculate the effective charge of the neutrinos we ultimately 
have to take the limit ${\bf k} \rightarrow 0$. Hence, eq.(\ref{pi5k0}) can be written as
\ber
&&\Pi^5_{3 0}(q_0 = 0, \vec q \rightarrow 0) \nonumber\\
&=& -\lim_{q_0=0,\vec{q}\rightarrow 0} \frac{i e^3 {\cal B}}{\pi} \int{d^2 p_\parallel \over{(2\pi)^2}} \eta_+(p_0)
     \int^{\infty}_{-\infty} ds\, e^{is(p^2_\parallel - m^2) - \varepsilon |s|}\nonumber\\
&\times& \int^{\infty}_0 ds' e^{is'(p'^2_\parallel - m^2) - \varepsilon |s'|} \left((q\cdot p)_\parallel - 2 p^2_0 \right)\nonumber \\
&=& \lim_{k_0=0,\vec{q}\rightarrow 0} 2 e^3 {\cal B} \int{d^2 p_\parallel \over{(2\pi)^2}} \eta_+(p_0)
    \frac{(p^2_\parallel - m^2)}{(p'^2_\parallel - m^2) + i\varepsilon } \nonumber \\
&\times& \left((q\cdot p)_\parallel - 2 p^2_0\right) \,,
\label{pi530}
\eer
where we have used the following relation,
\ber
\int^{\infty}_{-\infty} ds \, e^{is \, (p^2_\parallel - m^2) - \varepsilon \, |s|}
   \, \int^{\infty}_0 ds' e^{is' \, (p'^2_\parallel - m^2) - \varepsilon \, |s'|} \nonumber \\
= 2 \pi i \frac{\delta(p^2_\parallel - m^2)}{(p'^2_\parallel - m^2) + i\varepsilon} \,.
\eer
The expression for $\Pi^5_{3 0}$ given by Eq.(\ref{pi530}) contains two parts which can be integrated separately to obtain,
\ber
\lim_{q_0=0,\vec{q} \rightarrow 0} \int \frac{d^2 p_\parallel}{(2 \pi)^2}(2 \pi)
\frac{(p^2_\parallel - m^2)}{(p'^2_\parallel - m^2) + i\varepsilon} \eta_+(p_0)(q\cdot p)_\parallel \nonumber\\
= \frac{1}{2}\int \frac{dp}{2 \pi}\,\, \frac{\eta_+(E_p)}{E_p} \,,
\label{int1}
\eer
and,
\ber
\lim_{q_0=0,\vec{q}\rightarrow 0} \int \frac{d^2 p_\parallel}{(2 \pi)^2}(2 \pi) {p_0}^2
\frac{(p^2_\parallel - m^2)}{(p'^2_\parallel - m^2) + i\varepsilon} \eta_+(p_0)\nonumber\\
= \frac{1}{4} \int \frac{dp}{2 \pi}\,\, [\frac{\eta_+(E_p)}{E_p} - \beta \eta_+(E_p)] \,,
\label{int2}
\eer
where $E_p^2 = p^2 + m^2$. Therefore, we have,
\ber
\Pi^5_{3 0}(q_0 = 0, \vec q \rightarrow 0) = \frac{( e^3 {\cal B})}{2\pi} \beta \int \frac{dp}{2 \pi} \eta_+(E_p).  
\label{complexp}
\eer

Now it is evident that in the limit ($q_0 = 0, {\bf q} \rightarrow 0$) the dominant component of $\Pi^5_{\mu \nu}$
is $\Pi^5_{30}$. Therefore, the index $\mu$ in Eq.(\ref{nec2}) can only take the value 3. Incorporating this fact 
and after taking the trace, Eq.(\ref{nec2}) takes the following form:
\beq
e_{\rm eff} = - \frac{G_F}{\sqrt{2}e} \, g_{\rm A} \, (1 - \lambda) \, \Pi_{30}^5(q_0=0, {\bf q}\rightarrow 0) \cos \theta \,,
\label{nec3}
\eeq
where, $\theta$ is the angle between the magnetic field and the direction of the neutrino propagation. Therefore, in the limit 
of $m \geq \mu$ we obtain :
\ber
e^{\nu}_{\rm eff} &=& \frac{1}{8 \sqrt{2} \pi^3} \, e^2 \, G_F \, g_{\rm A}
                      \, (1 - \lambda) \, {\cal B} \, \cos \theta \nonumber \\
                      && \times \, \sum^{\infty}_{n=0} (-1)^n \, (1+n) \, \beta m \, K_1[(n+1) \beta m] \nonumber \\
                      && \times \, \cosh \{(n+1) \beta \mu\} \,,
\eer
where $K_n$ is the n-th order modified Bessel function of the second kind. It should be noted here that even though this
result is linear in the field strength ${\cal B}$, this result is exact to all odd powers of ${\cal B}$. In the zero frequency,
long wavelength limit it is only the term linear in ${\cal B}$ that survives. To get a feeling of the magnitude of the effective
charge of a neutrino in presence of a background magnetic field, we compare this with that in absence of a magnetic field. In
the limit of vanishing chemical potential, the ratio between the effective charge in a magnetised medium and in a simple
thermal medium turns out to be :
\beq
\frac {e^{\nu}_{eff}({\cal B})}{e^{\nu}_{eff}({\cal B}=0)}
= \frac{c_A}{4\pi^3} \Big( \frac{\cal B}{{\cal B}_c} \Big) (m \beta)^{3} K_1(m\beta) \cos \theta
\eeq
It can be seen from the equation above that the ratio is proportional to $\frac{\cal B}{{\cal B}_c}$. Since, in almost all
astrophysical situations encountered so far, this ratio is less than one, in the weak field limit the charge due to the
unmagnetized plasma is larger than in the case of a magnetised plasma. 

\section{Conclusion}
\label{concl}
 
To conclude, we note that only left-handed neutrinos acquire an effective charge. Since we have performed our calculation
for massless, standard-model neutrinos that should automatically come out of the theory. But recent observations indicate that
the neutrinos have mass (which allows for both left-handed and right-handed neutrinos). Our treatment can be modified for 
massive neutrinos following the method adopted in \cite{pal2} and we expect that the qualitative aspects of our result should 
remain the same.
 
More importantly, we notice that the presence of a magnetic field breaks the isotropy of space and introduces a preferential 
direction. As a consequence neutrinos propagating along the direction of the magnetic field acquire a positive charge whereas 
those propagating in the opposite direction acquire a negative charge. The net effect of this would then be the creation of a 
charge current along the direction of the field. Interestingly, neutrinos propagating in a direction perpendicular to the field 
would acquire zero effective charge. Whereas in an unmagnetized thermal medium neutrinos acquire an effective charge irrespective 
of their direction of propagation. Therefore for isotropic neutrino propagation no net current is generated in that case unlike 
in the presence of a magnetic field.
 
It is worth mentioning here that the generation of this charge current may play a significant role in the magnetic field 
generation of neutron stars being produced in supernova explosion. Another possible application could be to the neutrino 
wind driven instability responsible for blowing up the outer mantle of supernova as proposed by \cite{shkl}. The basic mechanism
of this instability can be understood as follows. Consider the collision of two plasma fluids. If we consider the motion of
one plasma with respect to the centre of mass of the other then the dispersion relation of the particles of the first would
depend on the relative velocity between the two plasma. Now this velocity dependent dispersion relation can give rise to
damping or instability of the plasma modes depending on the relative velocity between the two media. The growth rate of the 
plasmons, in such systems, has been estimated using formalism of finite temperature field theory \cite{bent,expl} 
as well as using plasma physics techniques \cite{shkl}. It is important to note that the finite temperature field theory 
techniques show the damping/growth to be proportional to $G^2_F$ whereas that calculated using the plasma physics techniques
show a scaling as $G_F$ \cite{shkl}.

In conclusion, we have calculated the effective charge of neutrinos in a weakly magnetized plasma, in the limit
$m > \mu$. It is observed that the neutrino charge acquires a direction dependence as a result of the presence
of an external magnetic field and it is also proportional to the magnitude
of the field strength present in the system.                                      
\section*{Acknowledgment}
We would like to thank N.~D. Hari Dass and Palash B. Pal for helpful discussions. We also thank Patrick Aurenche for going 
through the preliminary draft and drawing out attention to some important aspects. AKG would like to thank Laboratoire de 
Physique Theoretique, Annecy, France for supporting a visit where a part of this work was carried out.
\beb
\bi{raff} G.~G. Raffelt, {\em Stars as Laboratories for Fundamental Physics}, (University of Chicago Press, 1996)
\bi{pal2} J.~F. Nieves and P.~B. Pal, Phys. Rev. D {\bf 49}, 1398 (1994)
\bi{alth} T. Altherr and P. Salati, Nuc. Phys. B {\bf 421}, 662 (1994)
\bi{colp} J. Cooperstein, Phys. Rep. {\bf{163}}, 95 (1988); H.~A.~Bethe,  Rev. Mod. Phys. {\bf{62}},801 (1990);
          S. Bludman, Da Hsuan Feng, Th. Gaisser and S. Pittel, Phys. Rep. {\bf{256}}, 1 (1995) 
\bi{expl} R.~C. Duncan, S.~L. Shapiro and I. Wasserman, Astrophys. J. {\bf 309}, 141 (1986);
          S.~A.~Colgate and R.~H. White, Astrophys. J. {\bf 143}, 626 (1986); H.~A. Bethe and
          J.~R. Wilson ibid {\bf 295}, 14 (1985); M Ramp and H.~T. Janka, astro-ph/0005438
\bi{shkl} L.~O.~Silva et. al, Phys. Rev. Lett. {\bf 83}, 2703 (1999) and references therein
\bi{kouv} C. Kouveliotou, T. Strohmayer, K. Hurley, J. van Paradijs, M.~H. Finger, S. Dieters,
          P. Woods, C. Thompson, R.~T. Duncan, Astrophys. J. {\bf 581}, L103 (1999)
\bi{chan} G. Chanmugam, Ann. Rev. Astron. \& Astrophys. {\bf 30}, 143 (1992)
\bi{hari} L.~L. DeRaad Jr., K.~A. Milton and N.~D. Hari Dass, Phys. Rev. D {\bf 14}, 3326 (1976)
\bi{pal1} J.~C. D'Olivio, J.~F. Nieves and P.~B. Pal, Phys. Rev. D {\bf 40}, 3679 (1989)
\bi{iorf} A.~N. Ioannisian and G.~G. Raffelt, Phys. Rev. D {\bf 55}, 7038 (1997), hep-ph/9612285
\bi{hame} S.~J. Hardy and D.~B. Melrose, Publ. Astron. Soc. Aus. {\bf 13}, 144 (1996)
\bi{orae} V.~N. Oraevsky, V.~B. Semikoz and Ya.~A. Smorodinsky, JETP Lett. {\bf 43}, 709 (1986)
\bi{shai} R. Shaisultanov, Phys. Rev. D {\bf 62}, 113005 (2000), hep-th/0002079
\bi{gsh1} H. Gies and R. Shaisultanov, Phys. Rev. D {\bf 62}, 73003 (2000), hep-ph/0003144
\bi{gsh2} H. Gies and R. Shaisultanov, Phys. Lett. B {\bf 480}, 129 (2000), hep-ph/0009342
\bi{frd1} A.~K. Ganguly, S. Konar and P.~B. Pal, Phys. Rev. D {\bf 60}, 105014 (1999), hep-ph/9905206
\bi{schw} J. Schwinger, Phys. Rev. {\bf 82}, 664 (1951)
\bi{tsai} W.~Y. Tsai, Phys. Rev. D {\bf 10}, 1342 and 2699 (1974)
\bi{ditt} W. Dittrich, Phys. Rev. D {\bf 19}, 2385 (1979)
\bi{elmf} P. Elmfors, D. Grasso and G. Raffelt, Nucl. Phys. B {\bf 479}, 3 (1996)
\bi{bent} L. Bento, Phys. Rev. D {\bf 61}, 13004 (2000)     
\bi{expl} J.~F. Nieves, Phys. Rev. D61, 113008, 2000; for a similar conclusion using Relativistic Quantum Kinetic theory see: 
          H.~T.~Elze, T.~Kodama and R. Opher, Phys. Rev. D {\bf 63}, 13008 (2000)
 
\eeb
\end{document}